\documentclass[reprint, amsmath,amssymb, aip]{revtex4-2}
\setcitestyle{super}%,open={},comma}
\usepackage{amsthm,amsmath,amssymb}

\usepackage{graphicx}% Include figure files
\usepackage{xcolor}
\usepackage[utf8]{inputenc}
\usepackage{amsmath}
\usepackage{mathtools}
\usepackage[capitalise]{cleveref}
\usepackage{listings}
\usepackage{pgfplots}
\pgfplotsset{compat=1.7}
\usepackage{appendix}
\usepackage[ruled,lined]{algorithm2e}
\usetikzlibrary{external}
 \newcommand{\commentoutA}[1]{}

\makeatletter
\begin{document}
    
    \preprint{LA-UR-24-24032}
    
    %\title{A factorization algorithm for the inverse overlap matrix \\ using Tensor cores and mixed precision}
    \title{Efficient Mixed-Precision Matrix Factorization of the Inverse Overlap Matrix in Electronic Structure Calculations with AI-Hardware and GPUs}
    \author{Adela Habib}
    \email{ahabib@lanl.gov}
    \author{Joshua Finkelstein}
    \email{jdf@lanl.gov}
    \author{Anders M. N. Niklasson}
    \email{amn@lanl.gov}
    \affiliation{Theoretical Division, Los Alamos National Laboratory, Los Alamos, New Mexico 87545}

    \date{\today}
    
    \begin{abstract}
        In recent years, a new kind of accelerated hardware has gained popularity in the Artificial Intelligence (AI) and Machine Learning (ML) communities which enables extremely high-performance tensor contractions in reduced precision for deep neural network calculations.  In this article, we exploit Nvidia Tensor cores, a prototypical example of such AI/ML hardware, to develop a mixed precision approach for computing a dense matrix factorization of the inverse overlap matrix in electronic structure theory, $S^{-1}$. This factorization of $S^{-1}$, written as $ZZ^T=S^{-1}$, is used to transform the general matrix eigenvalue problem into a standard matrix eigenvalue problem. Here we present a mixed precision iterative refinement algorithm where $Z$ is given recursively using matrix-matrix multiplications and can be computed with high performance on Tensor cores. To understand the performance and accuracy of Tensor cores, comparisons are made to GPU-only implementations in single and double precision. Additionally, we propose a non-parametric stopping criteria which is robust in the face of lower precision floating point operations. The algorithm is particularly useful when we have a good initial guess to $Z$, for example, from previous time steps in quantum-mechanical molecular dynamics simulations or from a previous iteration in a geometry optimization. 
    \end{abstract}
    
    \maketitle

    \section{Introduction}
    The present-day boom in specialized hardware designed for artificial intelligence (AI) applications presents exciting opportunities for scientific computing. AI-specific hardware opens up new avenues for tackling some of the most computationally intensive problems with unprecedented speed and efficiency. However, this does not come without its challenges; existing solvers and algorithms often need to be adapted or replaced to effectively leverage these new hardware architectures. %nice!

    In this article we focus on the general quantum-mechanical eigenvalue problem in electronic structure theory that appears when we use a local, non-orthogonal, atom-centered basis set. We explore how to leverage AI-hardware for computing the congruence transformation of the generalized eigenvalue problem into a standard eigenvalue problem. We show how the transformation matrices, which are determined by a factorization of the inverse overlap matrix, can be generated recursively using generalized dense matrix-matrix multiplications. These multiplications are ideally suited for AI-hardware optimized for tensor contractions in convolutional deep neural networks. The technique is based on an iterative refinement algorithm for the factorization of the inverse overlap matrix, which was originally designed to achieve a computational complexity that scales linearly with the system size using numerically thresholded sparse matrix algebra.\cite{ANiklasson04b,ERubensson08c} Algorithms for how we thereafter can calculate the electronic structure using AI-hardware such as Tensor cores or Tensor Processing Units have been presented elsewhere.\cite{JFinkelstein21a,JFinkelstein21b,JFinkelstein22,RPederson22}

    Currently, AI-accelerators often utilize low-precision floating-point arithmetic to achieve maximum performance, potentially impacting accuracy and convergence. We will address these challenges with mixed precision formulations and provide parameter-free convergence conditions. \cite{akruchinina16,JFinkelstein21a} We then demonstrate the effectiveness of our approach using Nvidia Tensor cores on an Nvidia A100 GPU \cite{nvda-a100} for both a test overlap matrix that is well-conditioned, and a more challenging, ill-conditioned overlap matrix for a silver nanoparticle generated by GPAW, the projector-augmented wave and multi-basis electronic structure code.\cite{EnkovaaraGPAWSoftware2010,Mortensen_2024} Although applied to a specific architecture, i.e.\ Tensor cores, our approach is quite general and should also be transferable to other forms of AI-accelerated hardware.

    Several alternative methods for computing the congruence transformation to a standard eigenvalue problem are well-established, such as computing a Cholesky factorization of the overlap matrix or computing the inverse overlap matrix via Schulz iterations. \cite{GSchulz33} The Schulz matrix inversion works well with both sparse matrix algebra and AI hardware, \cite{RPederson22} though direct methods, such as Cholesky factorizations, do not naturally map onto Tensor core hardware. We do not address these alternative techniques here.

    After first presenting some background on the generalized quantum-mechanical eigenvalue problem in electronic structure theory, we discuss the congruence transformation for the generalized eigenvalue problem using the factorization of the inverse overlap matrix. We then discuss the new opportunities and challenges associated with new AI-hardware, such as Tensor cores in Section \ref{sec:background}. Next, we propose to use an iterative refinement algorithm for a factorization of the inverse overlap matrix with a mixed precision implementation in Section \ref{sec:algorithm} and also discuss adaptive approaches with an adjustable floating point precision. Thereafter, in Section \ref{sec:stopping}, we present a parameter-free convergence criteria which provides conditions on how to terminate the mixed-precision iterative calculation. Numerical results are shown in Section \ref{sec:numerics} for the silver nanoparticle and synthetic test case that compares accuracy and performance of our iterative technique when using Tensor cores with various precisions. A comparison is also provided to explicit construction of $S^{-1/2}$ with diagonalization using Nvidia's cuSolver library. We conclude with a brief summary. 

    \section{Background}\label{sec:background}

    The matrix factorization algorithm for the inverse overlap matrix that we will design for AI-hardware is quite general. However, our main motivation and target problem that we will focus on is the quantum-mechanical eigenvalue problem that occurs in electronic structure theory, for example, in Hartree-Fock and density functional theory (DFT), \cite{croothaan51,PHohenberg64,WKohn65,RParr89, RDreizler90} or in various semi-empirical methods,\cite{MElstner98,MFinnis98,MDewar77,JStewart13,PDral19,WMalone20,BHourahine20,ZGuoqing20,CBannwarth20} when we use atom-centered, non-orthogonal, basis sets. To motivate and explain this target problem, we present the generalized quantum-mechanical eigenvalue equation as it appears in Kohn-Sham DFT. 

    \subsection{Kohn-Sham density functional theory}

    In Hohenberg-Kohn density functional theory
    \cite{PHohenberg64,RParr89,RDreizler90} the ground state electron density, $\rho_{\rm min}(\bf r)$, of an atomistic system is given from a constrained minimization over all physically relevant electron densities, $\rho({\bf r})$, where
    \begin{equation}\label{rho_min}
       {\displaystyle \rho_{\rm min}(\bf r) = \arg \min_{\rho} \left\{ E(\bf R,\rho) \left \vert ~\int \rho(\bf r) d\bf r = N_e \right. \right\} },
    \end{equation}
    with an energy functional $E$ defined by, 
    \begin{equation}\label{EDFT}
       {\displaystyle  E(\bf R, \rho ) = F[\rho] + \int \rho(\bf r) v_{\rm ext}(\bf R,\bf r ) d\bf r}.
    \end{equation}
    Here $F[\rho]$ is a system-independent universal functional, $v_{\rm ext}(\bf R,\bf r)$ is the external potential from the atomic nuclei, and $N_e$ is the number of electrons. From the ground state electron density, a number of physical observables can then be calculated. 

    Kohn-Sham (KS) density functional theory \cite{WKohn65,RParr89,RDreizler90} is the most common way to represent the density and the universal functional, $F[\rho]$. The electron density is assumed to be a sum over $N_e/2$ single-particle orbitals,
    \begin{align}
    \begin{split}\label{charge}
    \rho({\bf r}) &= 2\sum_k f_k \vert \psi_k({\bf r})\vert^2, \\
    &~~ \text{such that}  \int \vert \psi_k({\bf r})\vert^2 d{\bf r} = 1, ~~ \forall k \;.
    \end{split}
    \end{align}
    The factor 2 is included under the assumption that each occupied orbital consists of two electrons (in spin up and down states) and the $f_k$ are the occupation factors ($f_k = 1$ for the occupied orbitals and $f_k = 0$  for the unoccupied ones). 
    To represent the single-particle orbitals, $\{\psi_k\}$, and the density, $\rho({\bf r})$, we use an approximate finite basis set expansion, where
    \begin{equation}
        \psi_k({\bf r}) = \sum_i^N c_i^{(k)} \varphi_i({\bf r}).
    \end{equation}
    These basis functions, $\{\varphi_i\}_{i=1}^N$, can be chosen, for example, as approximate atom-centered, non-orthogonal, local atomic-orbitals. In this representation, the corresponding Kohn-Sham minimization in Eq.\ (\ref{rho_min}), is given in terms of a generalized matrix eigenvalue equation,
    \begin{equation}\label{KS-Eig}
        HC = SC{\boldsymbol \epsilon},
    \end{equation} 
    with the Kohn-Sham Hamiltonian matrix, 
    \begin{equation}\label{KS_Matrix}
        H_{ij} = \int \varphi_i^*({\bf r}) \left( -\frac{\hbar^2}{2m} \nabla^2 + V_{\rm KS}[{\bf R}, \rho]({\bf r}) \right) \varphi_j({\bf r}) \; d{\bf r},
    \end{equation}
    the overlap matrix,
    \begin{equation}\label{Overlap}
        S_{ij} = \int \varphi_i^*({\bf r}) \varphi_j ({\bf r})\; d{\bf r},
    \end{equation}
    and the eigenvector coefficient and eigenvalue matrix,
    \begin{equation}
        C_{ij} = c_i^{(j)}, ~~ {\boldsymbol \epsilon}_{ij} = \delta_{ij}\epsilon_i\; .
    \end{equation}    
    The function $V_{\rm KS}[{\bf R},\rho]({\bf r})$ is the $\rho$-dependent Kohn-Sham effective single-particle potential that depends on the external potential and how we choose to approximate $F[\rho]$. $-\hbar^2 \nabla^2/(2m)$ is the kinetic energy term of the Kohn-Sham Hamiltonian.\cite{WKohn65,RParr89,RDreizler90}
    From the eigenvector, $\{c_i^{(j)}\}$, the electron density is given by 
    \begin{equation}
        \rho({\bf r}) = \sum_{i,j,k}f(\varepsilon_k) c_i^{(k)}c_j^{(k)}\varphi_i^*({\bf r})\varphi_j({\bf r}).
    \end{equation}

    Because the Kohn-Sham Hamiltonian depends on the electron density through the Kohn-Sham potential, $V_{\rm KS}[{\bf R},\rho]$, the ground state optimization has to be performed iteratively, and the generalized eigenvalue problem in Eq.\ (\ref{KS-Eig}) has to be solved for each new updated electron density until a converged self-consistent ground-state solution is found.

    To repeatedly solve the generalized non-linear eigenvalue problem in Eq.\ (\ref{KS-Eig}), we first need to transform the equation into a standard form, where the overlap matrix is avoided. This can be achieved with a congruence transformation.\cite{GGolub96} If $Z$ is a matrix determined by
    \begin{equation} \label{Def_Z}
        Z^TSZ = I\;, {\rm i.e.}~ZZ^T = S^{-1}, 
    \end{equation}
    we can transform the generalized eigenvalue problem into the standard form,
    \begin{equation}\label{eq:standard_form}
        H^\perp C^\perp = C^\perp {\boldsymbol \epsilon},
    \end{equation}
    where $H^\perp = Z^THZ$ and $C^\perp = Z^{-1}C$. Equation (\ref{eq:standard_form}) is then solved, for example, using a diagonalization algorithm based on, for example, QR iterations \cite{Anderson92,GGolub96} or divide-and-conquer.\cite{JCuppen80,cuSOLVER}

    \subsection{Tensor cores and AI Hardware}

    A factorization of the inverse overlap matrix, as defined by Eq.\ (\ref{Def_Z}), can be performed using a diagonalization of the overlap matrix, $S$, from which we then can calculate, $Z = S^{-1/2}$. However, diagonalization algorithms are in general not suitable for AI-hardware such as Tensor cores, and are also not particularly efficient for smaller matrix sizes on an ordinary GPU.\cite{JFattebert24} For example, QR iterations are highly non-local and hard to parallelize efficiently. Similarly, with divide-and-conquer approaches, tridiagonalization makes use of memory-bound BLAS Level 2 (i.e. matrix-vector) operations.\cite{WYu21} To be able to leverage AI-tailored hardware architectures, we need algorithms that naturally map onto the computational structure of a deep neural network where the computational kernel is focused on tensor contractions. Recursive matrix-function expansion methods, where the computational cost is dominated by matrix-matrix operations, are therefore particularly well suited. 
    
    In this article, we focus on an iterative refinement approach that is based on recursive expansions for the factorization of the inverse overlap matrix. Originally, this set of algorithms was designed for numerically thresholded sparse matrix algebra to achieve linear scaling complexity for sufficiently large and sparse matrix problems. \cite{ANiklasson04b,Branislav07,ERubensson08c,CNegre16} Here we use the algorithm for dense matrix algebra with mixed precision floating point operations. This means that the computational cost for the matrix-matrix multiplications scale cubicly with the system size. Although this may appear inefficient compared to using numerically thresholded sparse matrix algebra, which scales only linearly with the system size,\cite{CNegre16} thanks to the exceptional performance of AI-hardware using dense matrix algebra, the advantage of sparse linear scaling schemes appears only for very large systems. It is in this intermediate regime (before linear scaling schemes become faster), with overlap matrices up to a few thousand basis-functions, where our methodology is of significance. For very large systems, e.g.\ for problems including hundreds of thousands or millions of atoms, linear-scaling sparse matrix algebra running even in serial on a single CPU is likely much faster than any dense matrix algebra implementation running in parallel on a distributed accelerated hardware platform.\cite{RPederson22} However, using graph theory, such large problems can often be broken down into smaller overlapping sub problems that can be solved in parallel with dense matrix algebra. \cite{ANiklasson16,Djidjev16,MLAss18,Djidjev19,MLass20,RSchade22,CNegre23} The algorithms presented in this paper can also become highly efficient for very large problems if they are combined with these graph-based approaches to electronic structure calculations.  

    A single Tensor core unit performs a block fused-multiply add operation that computes $D=A \times B + C$, where the specific matrix dimensions depend on the architecture. For Tensor cores on an Nvidia A100 GPU, $A$ is a $4 \times 8$ matrix, $B$ is an $8 \times 8$ matrix and $C$ is a $4\times 8$ matrix \cite{nvda-a100} while, for example on Nvidia V100 Tensor cores, $A$, $B$, and $C$ are all $4 \times 4$ matrices.\cite{nvda-v100} The CUDA language has the ability to perform these basic Tensor core operations at a low level through the {\tt wmma} API functionality (which stands for warp-level matrix multiply accumulate), but the easiest way to use Tensor cores for high-performance is through libraries such as Nvidia's cuBlas, which is readily available and strongly supported by Nvidia. This is the approach we take, although other implementations of high-performance GEMMs (General Matrix Multiply) that utilize Tensor cores are available.\cite{cutlass,HOotomo22,MFasi2023}

    \section{Inverse Overlap Matrix Factorization Algorithm}\label{sec:algorithm}
    
    The iterative refinement algorithm for the inverse factorization of the overlap matrix as introduced in Ref.~\citenum{ANiklasson04b} assumes an initial guess $Z_0$ that is sufficiently close to a factor of $S^{-1}$ such that
    \begin{equation}
        X_0 = Z^T_0SZ_0 \approx I.
    \label{eq:zero}
    \end{equation}
    To refine $Z_0$ and find a more accurate $S^{-1}$ factor, the algorithm
    takes the following form,
    \begin{align}
    \begin{split}\label{eq:np1}
        Z_{n+1} &= Z_n\bigg(\sum^{m}_{k=0} a_k X^{k}_n\bigg)\;,\\
        X_{n+1} &= Z^T_{n+1}SZ_{n+1},
    \end{split}
    \end{align}
    which is iterated for integers $n \ge 0$ until convergence. The iterated $X_n$ converges as
    \begin{equation}
        \lim_{n\to \infty} X_n =\lim_{n\to\infty} Z^T_nSZ_n = I\;, ~ \text{if} ~ ||X_0 - I||_2 < 1\;,
    \end{equation}
    with $Z_n \to Z$, an inverse overlap factor. The scheme for $m = 2$ is given in Algorithm \ref{algorithm:inv-overlap}. The coefficients, $a_k$, in the polynomial in Eq.~(\ref{eq:np1}) are determined from the condition that the error terms, 
    \begin{equation}
        \delta_n^\nu = (X_n - I)^\nu\;,
    \label{eq:errorConv}
    \end{equation}
    should vanish up to the highest possible order, $ \nu = \nu_{\text max}$, after each iteration in Eq.\ (\ref{eq:np1}). The error $X_n-I$ is therefore expected to decay as $\mathcal{O}((X_0-I)^{(\nu_\text{max})^n})$ for iteration $n$. Reference~\citenum{ANiklasson04b} gives a table with the coefficients $a_k$ for polynomials of order $m= 2,3,4, 5$. For the tests in this article, we use $m=2$ so that $\nu_{\text max} = 3$. Readers can refer to Ref.~\citenum{ANiklasson04b} for more details of the algorithm and convergence tests. The single and double precision implementations of the algorithm in Eq.~(\ref{eq:np1}) will be denoted {\bf single} and {\bf double}, respectively, in the text. 

    \begin{algorithm}
        Z = initial guess \\ 
        $\varepsilon = $ some small number \\
       {\rm iter} = 0\\
        $X = Z^TSZ$ \\
        \For{${\rm iter} < {\rm Nmax}$}{
            {\rm Err} = $\|X-I\|_F$\\
            \If {${\rm Err} < \varepsilon$} {
                break\\
            }
            {\rm iter} = {\rm iter}+1\\
            $Z = Z(a_0I + a_1X + a_2X^2)$ \\
            $X = Z^TSZ$ \\
        }
        \caption{The iterative inverse overlap matrix factorization algorithm as presented in Ref.~\citenum{ANiklasson04b} with $m=2$.}\label{algorithm:inv-overlap}
    \end{algorithm}

    As an alternative, and equivalent form, we could replace $X_n$ by $\delta_n = X_n-I$ in the polynomial expansion, where 
    \begin{equation}
        X_{n+1} = Z_n \left(\sum_{k=0}^m c_k \delta_n^k \right).
    \end{equation}
    In this case the polynomial coefficients, $\{c_k\}$, can be determined more easily, e.g.\ see Refs.\ \citenum{ERubensson08c,EHRubensson20,Branislav07}.

    \subsection{Matrix splitting}\label{sec:splitting}
    The main contribution to the computational cost of the iterative factorization algorithm are the dense matrix-matrix multiplications, which scale cubicly with the size of the matrix. In order to achieve maximum performance with Tensor cores, these matrix multiplications need to be performed using a low precision format, where input matrices are represented in half-precision and products are accumulated in single-precision. In applications such as electronic structure calculations or quantum-based molecular dynamics simulations however, this format does not typically provide enough accuracy to be useful. A simple, yet effective technique for addressing this that we have used in the past \cite{JFinkelstein21a,JFinkelstein21b,JFinkelstein22} is to represent a single-precision matrix $A$ as a sum of two half-precision matrices, where 
    \begin{equation}\label{dualRep}
        A \approx A^{(h)} + A^{(l)}
    \end{equation}
    with the matrix splitting
    \begin{equation}
        \begin{array}{ll}
            A^{(h)} &= {\rm FP16}[A] \\
            A^{(l)} &= {\rm FP16}[A-{\rm FP32}[A^{(h)}]].
        \end{array} \label{eq:splitStyle}
    \end{equation}
    Here, and in what follows, half, single and double precision floating-point representations are denoted using the notation FP16, FP32 and FP64, respectively. In our notation, $h$ represents the ``high" and most significant part of the floating point mantissa of $A$ and $l$ denotes the ``low" part. Using this combined half-precision representation in Eq.\ (\ref{dualRep}) the maximal componentwise error of $A$ is $\Delta = \mathcal{O}({u}_{{\rm half}}^2)$, where ${u}_{{\rm half}}=2^{-11}$ is the unit round-off for the half precision floating-point representation with 10 bits in the mantissa.\cite{MFasi2023} A bound on the error of the combined half-precision representation can be explicitly computed using the standard model of floating point arithmetic,\cite{MFasi2023,NHigham02} by observing that the $A^{(h)}$ and $A^{(l)}$ can be written as
    \begin{align}
        A^{(h)} &= A\circ({\bf 1} + \Delta_h) \\
        A^{(l)} &= (A-A^{(h)})\circ({\bf 1} + \Delta_l) 
    \end{align}
    for some componentwise error matrices $\Delta_h, \Delta_l\in \mathbb{R}^{N \times N}$, where $|(\Delta_h)_{ij}| \le {u}_{{\rm half}}$,  and $|(\Delta_l)_{ij}| \le {u}_{{\rm half}}$, for $1\le i,j \le N$. The $\circ$ notation represents the componentwise (Hadamard) product for matrices and ${\bf 1}$ is an $N\times N$ matrix of ones. Using this notation, the sum is then  
    \begin{align*}
            A^{(h)} + A^{(l)} & = A\circ({\bf 1} + \Delta_h) + (A-A^{(h)})\circ ({\bf 1} + \Delta_l) \\
                          & = A\circ({\bf 1} + \Delta_h) \\
                          & \qquad +  (A-A\circ({\bf 1} + \Delta_1))\circ({\bf 1} + \Delta_l) \\
                          & = A + A\circ\Delta_h - A\circ\Delta_h\circ({\bf 1} + \Delta_l) \\
                          & = A - A\circ\Delta_h\circ\Delta_l \;,
    \end{align*}
    so that $A = A^{(h)} + A^{(l)} + E$, where $E$ is an error term with $|E_{ij}|\le {u}_{\rm half}^2|A_{ij}| = 2^{-22}|A_{ij}|$. The sum $A^{(h)} + A^{(l)}$ is thus fairly close to a single precision floating-point representation of $A$, which has 23 bits in the mantissa with ${u}_{\rm single} = 2^{-24}$.
    
    \subsection{Adapting the inverse overlap algorithm for Tensor cores: the $(h)(h)+(h)(l)+(l)(h)$ scheme}
    We take two different approaches to implementing Algorithm~\ref{algorithm:inv-overlap} for $m = 2$ on Tensor cores. First, we apply the splitting techniques of Sec.~\ref{sec:splitting} to the matrices in the iterative inverse overlap algorithm. The matrices $Z$ and $S$ are stored in single precision and are then split into their FP16 representations as in Eqs.\ (\ref{dualRep}) and (\ref{eq:splitStyle}), i.e.\
    \begin{equation}
        \begin{array}{ll}
            S   &= S^{(h)} + S^{(l)}, \\
            Z_n &= Z_n^{(h)} + Z_n^{(l)}.
        \end{array}
    \end{equation}

    The scheme in Eq.~(\ref{eq:np1}) requires that triple matrix products are calculated. To compute $X_n = Z^T_nSZ_n$, we first apply the matrix splitting approach to
    $SZ_n$, 
    \begin{align}
        A &= SZ_n \\
          &= (S^{(h)} + S^{(l)})(Z_n^{(h)} + Z_n^{(l)}) \\
          &= S^{(h)}Z_n^{(h)} + S^{(h)}Z_n^{(l)} + S^{(l)}Z_n^{(h)} + S^{(l)}Z_n^{(l)}\\
          &\approx S^{(h)}Z_n^{(h)} + S^{(h)}Z_n^{(l)} + S^{(l)}Z_n^{(h)} \;.
     \label{eq:SZn}
    \end{align}
    where in the last line we discard the $(l)(l)$ term as they are expected to be small contributions to the overall error. This is consistent with our previous results, \cite{JFinkelstein21a,JFinkelstein21b} and is also theoretically justified.\cite{MFasi2023} The calculated $A$ is then split and represented by the combined half-precision representation, i.e.\ $A \rightarrow A^{(h)} + A^{(l)}$. We then compute $X_n$ as, 
    \begin{align}
        X_n &= Z^T_nSZ_n\\
            &= (Z_n^{(h)} + Z_n^{(l)})^T(A^{(h)} + A^{(l)})\\
            &= (Z_n^{(h)})^TA^{(h)} + (Z_n^{(l)})^TA^{(h)} \\ \nonumber
            & \hspace{2cm} + (Z_n^{(h)})^TA^{(l)} + (Z_n^{(l)})^TA^{(l)} \\
            &\approx (Z_n^{(h)})^TA^{(h)} + (Z_n^{(l)})^TA^{(h)} + (Z_n^{(h)})^TA^{(l)} \;.
    \label{eq:Xn}
    \end{align} 
    The calculated $X_n$ is then split and represented by the combined half-precision representation, i.e.\ $X_n \rightarrow X_n^{(h)} + X_n^{(l)}$.
    Lastly, to compute $Z_{n+1}$ in Eq.~(\ref{eq:np1}), the matrix product $Z_n (a_0I + a_1 X_n + b_2X_n^2)$ has to be evaluated. Computing the matrix square is done through 
    \begin{align}
        X^2_n & = (X_n^{(h)} + X_n^{(l)}) (X_n^{(h)} + X_n^{(l)})\\
              %& = X_n^{(0)}X_n^{(0)}+ X_n^{(0)}X_n^{(1)} \nonumber \\
              %& \hspace{2cm} + X_n^{(1)}X_n^{(0)} + X_n^{(1)}X_n^{(1)}\\
              & = X_n^{(h)}X_n^{(h)}+ X_n^{(h)}X_n^{(l)} \\ \nonumber
              & \hspace{2cm} +(X_n^{(h)}X_n^{(l)})^T + X_n^{(l)}X_n^{(l)}\\
              & \approx X_n^{(h)}X_n^{(h)}+ X_n^{(h)}X_n^{(l)} +(X_n^{(h)}X_n^{(l)})^T \;.
     \label{eq:Xsq}
    \end{align}
    since $X_n^{(l)}X_n^{(h)} = (X_n^{(h)}X_n^{(l)})^T$ by symmetry, which reduces the number of multiplications needed to evaluate the square from three to two. 
    
    The decomposition of each high precision multiplication into three lower precision multiplications gives the impression that we actually raise the computational cost, since the total number of matrix-matrix multiplications increases to 12 from four in one iteration of the algorithm. However because of the more than order-of-magnitude speedup of Tensor cores over ordinary dense GPU matrix multiplications, an overall speedup can still be achieved. This implementation of the inverse overlap factorization algorithm will be called the $(h)(h)+(h)(l)+(l)(h)$ scheme.

    \begin{widetext}
    \begin{table}[]
    \begin{minipage}{\textwidth} 
        \centering
        \begin{tabular}{p{0.15\linewidth}|p{0.45\linewidth}|p{0.27\linewidth}}
            Name of Method & Description & cuBlas API call\\
            \hline
            {\bf double} & All multiplications in double precision, with results accumulated in double precision. & {\tt cublasDgemm} \\
            {\bf single} & All multiplications in single precision, with results accumulated in single precision. & {\tt cublasSgemm} \\
            $(h)(h)$ & All matrix multiplications take half-precision representation of input and accumulate the results in single precision. & {\tt cublasGemmEx} \newline with  {\tt CUBLAS\_COMPUTE\_32F\_FAST\_16F}\\
            $(h)(h){\bf half}$ & All matrix multiplications take half-precision representation of input and accumulate the results in half precision. & {\tt cublasGemmEx} \newline with  {\tt CUBLAS\_COMPUTE\_16F} \\
            $(h)(h)+{\bf refi}$ & All matrix multiplications take half precision representation of the inputs and accumulate in single precision. A final single precision refinement step is performed after convergence criteria engages. &{\tt cublasGemmEx} \newline with  {\tt CUBLAS\_COMPUTE\_32F\_FAST\_16F} and {\tt CUBLAS\_COMPUTE\_32F} \\
            $(h)(h)+(h)(l)+(l)(h)$ & Each matrix is split into three parts using the scheme described in Sec.~\ref{sec:splitting} and multiplications are carried out with half precision input and single precision accumulation. No refinement step is used. & {\tt cublasGemmEx} \newline with {\tt CUBLAS\_COMPUTE\_32F\_FAST\_16F}\\
            $(h)(h)+(h)(l)+(l)(h)+{\bf refi}$ & Same as $(h)(h)+(h)(l)+(l)(h)$ with an added refinement step in single or double precision.& {\tt cublasGemmEx} \newline with  \newline  {\tt CUBLAS\_COMPUTE\_32F\_FAST\_16F} and {\tt CUBLAS\_COMPUTE\_32F}  or  \newline {\tt CUBLAS\_COMPUTE\_64F} \\
            \hline
        \end{tabular}
        \caption{This table describes the various versions of the iterative refinement algorithm in Eq.~(\ref{eq:np1}) and the precisions used for each version. Also included are the cuBlas library API calls and flags used for each method in our implementation.}
        \label{tab:methods}
    \end{minipage}
    \end{table}
    \end{widetext}

    \subsection{Hybrid $(h)(h)$ and  $(h)(h)+(h)(l)+(l)(h)$ schemes with refinement}
    
    The large number of matrix multiplications required for the splitting technique reduces the benefit of the accelerated hardware. An attractive alternative is to only compute the $(h)(h)$ multiplications in the above Equations (\ref{eq:SZn}) to (\ref{eq:Xsq}), which we call the $(h)(h)$ scheme. This minimizes the number of multiplications needed per iteration, going from 12 to just 4, and is equivalent to naively running the algorithm in half-precision (with FP32 accumulation in the matrix multiplication dot products). Although this may not be accurate enough to obtain a well-converged solution, we can achieve sufficient accuracy through additional refinement iterations in higher precision. That is, we run the algorithm using the $(h)(h)$ scheme and then do a final one or two iterations in single or double precision, taking advantage of the cubic convergence of $X_n$ (see Section \ref{sec:stopping}). The iterative method using this adaptive precision scheme will be called $(h)(h)+{\rm \bf refi}$.  

    For less well-conditioned overlap matrices, $(h)(h)$ may not be accurate enough to get a converged solution, or may take many iterations to converge. In this case we may add an additional refinement step to the more accurate $(h)(h)+(h)(l)+(l)(h)$ method, either in single or double precision. We call this adaptive scheme $(h)(h)+(h)(l)+(l)(h)+{\bf refi}$.
    
    \subsection{Implementation}
    The cuBLAS library provides matrix-matrix multiplication routines that execute on Tensor cores when proper flags are set. For mixed-precision multiplication, the $\texttt{cublasGemmEx}$ routine is suitable as it allows the specification of input and output data types. For single or double precision multiplications, we use the API calls {\tt cublasSgemm} or {\tt cublasDgemm}, respectively.  Several flags are provided to users that, according to the documentation, will run on Tensor cores ``when possible". To actually check whether Tensor cores are used, we either run a profiler or calculate the number of FLOPs. Properly setting these flags is thus important for achieving the desired accuracy and computational speed. In our Tensor core implementations, we specifically use \texttt{CUBLAS\textunderscore COMPUTE\textunderscore 32F\textunderscore FAST\textunderscore 16F} along with the splitting method discussed above. All methods described in this section along with how they are implemented are summarized in Table \ref{tab:methods}.
    
    \section{Practical Algorithms}\label{sec:stopping}
    Convergence is reached when $\|X_n-I\|$ is ``sufficiently'' small, which requires determination of a stopping parameter value. When working with mixed precision however, and switching between half and single precision for example, choosing a ``good" parameter value isn't immediately straightforward. The numerical noise level can be significant and the best possible convergence can differ depending on the condition number of the overlap matrices. The convergence problem may appear to be a small detail, but it is of great importance for practically useful algorithms. Here we avoid the problem of finding a convergence parameter value by designing a parameter-free stopping condition. The idea is to formulate an analytic convergence ratio between iterations that is satisfied in exact arithmetic, and then terminating the calculation when this analytical bound is no longer true. \cite{akruchinina16,JFinkelstein21a} In this sense, all available numerical precision has been exhausted and further iterations do not provide additional benefit.

    \subsection{Convergence condition}
    We base our convergence criteria on the decaying nature of the symmetric error $\delta^\nu = (X_n-I)^\nu$ after each iteration $n$ for a given polynomial order in Eq.~(\ref{eq:np1}). In our analysis below we don't specify the matrix norm but assume that it is submultiplicative (or consistent), i.e. for matrices $A$ and $B$, $\|AB\| \le \|A\|\|B\|$. We also only consider the case of $m=2$ in Eq.~(\ref{eq:np1}), and make an ansatz for $X_n$, $n\ge 0$, where,
    \begin{equation}
        \begin{array}{ll}
            X_n = I + \delta_n  \;. \\
        \end{array}
    \label{eq:X0}
    \end{equation}
    for some  small error matrix $\delta_n$.
    The next iterate, $X_{n+1}$, then has the form
    \begin{equation}
        \begin{array}{ll}
            X_{n+1} = I + a\delta_n + b\delta_n^2 + c\delta^3_n + d\delta_n^4 +  e\delta_n^5\;, & 
        \end{array}
    \label{eq:Xp1}
    \end{equation}
    for some coefficients $a,b,c,d,e$. By solving $X_{n+1} = Z_{n+1}^TSZ_{n+1}$ and requiring at most a cubic contribution, i.e. $a=b=0$, (which is how the coefficients $\{a_k\}$ are determined for in Algorithm \ \ref{algorithm:inv-overlap} for $m = 2$) the remaining coefficients can be determined and are $c$ = 5/8, $d$ = -15/64, and $e$ = 9/64 (see Appendix A for full calculation). The norm of $X_{n+1} - I$ is then
    \begin{align}\label{eq:norm2}
        ||X_{n+1}-I|| &= ||c\delta^3_n + d \delta_n^4 + e \delta_n^5\| \\
        & \le |c|\|\delta_n\|^3 + |d| \|\delta_n\|^4 + |e| \|\delta_n\|^5, 
    \end{align}
    which gives us the upper-bound on the ratio,
    \begin{equation}\label{eq:norm condition}
        \frac{\|X_{n+1}-I\|}{\||X_n-I\|^3} \leq |c| + |d|\|\delta_n\| + |e|\|\delta_n\|^2,
    \end{equation}
    where we used the fact that $\|\delta_n\|^3 = \|X_n-I\|^3$.

    Close to convergence, after repeated iterations ($n > 0$), the error, $\|\delta_n\|$, is small. If we assume that $\|\delta_n\| < 1$ is a sufficient condition for convergence (which is straightforward to check separately), the right-hand side in Eq.~(\ref{eq:norm condition}) will be less than unity. Our ratio of the errors between successive iterations then becomes bounded by
    \begin{equation}
        \frac{\|X_{n+1}-I\|}{\|X_{n}-I\|^3} < 1 \;, \text{ if } |\|X_n-I\|<1\;, 
    \label{eq:stopL2}
    \end{equation} 
    for general $n\ge 0$. 
    
    The convergence ratio inequality in Eq.\ (\ref{eq:stopL2}) can easily be used as condition for convergence. The bound of the error ratio between iterations in Eq.\ (\ref{eq:stopL2}) always holds in infinite precision. However, because of the finite precision floating point representations, the inequality eventually breaks down. At this point, there is no reason to continue the iterations, because the best possible convergence has already been reached. In this way we use the convergence ratio inequality in Eq.\ (\ref{eq:stopL2}), or Eq.\ (\ref{eq:norm condition}), as an efficient convergence condition and define convergence to be achieved once this inequality is broken. Also noteworthy is the fact that this bound proves cubic convergence of the iterative scheme.
    
    The parameter-free convergence condition requires the calculation of some submultiplicative (or consistent) matrix norm.  Calculation of 2-norms, $\| \cdot \|_2$, are usually expensive or cumbersome as they require determining the spectral radius. As an alternative we could use the 1-norm or infinity norm \cite{GGolub96} to estimate the convergence condition in Eq.\ (\ref{eq:stopL2}). However, the Frobenius matrix norm, $\| \cdot \|_{\rm F}$, is straightforward to compute and will be used in our examples throughout the text for the convergence condition, unless otherwise stated. The inverse overlap factorization algorithm with an implementation of the stopping condition is shown below in Algorithm \ref{algorithm:inv-overlap-wstopping}.
    
    \begin{algorithm}     
        Z = initial guess \\ 
        iter = 0 \\
        $X = Z^TSZ$ \\
        \For{${\rm iter}<${\rm Nmax}}{
            %\tcp{Check error}
            Err[iter] = $\|X-I\|_{\rm F}$ \\
            \If {${\rm iter} > 0 \text{ \rm \bf and }  {\rm Err [iter]} > ({\rm Err [iter-1]})^3$} {
                \rm break \\
            }
            iter = iter + 1 \\
            $Z = Z(a_0I + a_1X + a_2X^2)$ \\
            $X = Z^TSZ$ \\
        }
        \caption{Iterative inverse overlap matrix factorization algorithm with parameter-free stopping criterion.}\label{algorithm:inv-overlap-wstopping}
    \end{algorithm}
    
    \subsection{Refinement step for higher precision} \label{sec:refinement}
    With the parameter-free convergence criteria in Eq.\ (\ref{eq:stopL2}), we can determine when the inverse factorization algorithm has converged to the best possible result for the given precision. At this point we introduce the refinement step, where an additional iteration is performed in an enhanced precision (e.g.\ single or double) for a particular factorization scheme. We denote this new scheme by appending a ``$+{\bf refi}$" to method name. For example, if we add a refinement step to $(h)(h)$, the new scheme becomes $(h)(h)+{\bf refi}$. Implementation of a refinement to $(h)(h)$ consists of the steps:  
    \begin{enumerate}
        \item Converge at iteration $n$ with the $(h)(h)$ method until the convergence criteria in Eq.~(\ref{eq:stopL2}) is fulfilled.
        \item Convert $Z_n$ into enhanced precision (e.g.\ single or double) and recompute $X_n=Z_n^T S Z_n$ in the enhanced precision.
        \item Compute $Z_{n+1}$ in enhanced precision.
    \end{enumerate}
    Psuedocode is shown in Algorithm \ref{algorithm:inv-overlap-wstopping-refine}. The same procedure can be applied to the $(h)(h)+(h)(l)+(l)(h)$ scheme with an additional enhanced precision refinement, which gives us the $(h)(h)+(h)(l)+(l)(h)+{\bf refi}$ scheme.
    \begin{algorithm}     
        Z = initial guess \\ 
        iter = 0 \\
        $X = Z^TSZ$ \\
        \For{${\rm iter}<{\rm Nmax}$}{
            %\tcp{Check error}
            Err[iter] = $\|X-I\|_{\rm F}$ \\
            \If {${\rm iter} > 0 \text{ \rm \bf and }  {\rm Err [iter]} > ({\rm Err [iter-1]})^3$} {
                \rm break \\
            }
            iter = iter + 1 \\
            $Z = Z(a_0I + a_1X + a_2X^2)$ \\
            $X = Z^TSZ$ \\
        }
        \tcp{Refinement step} 
        $Z$ = FP32[$Z$] or FP64[$Z$] \\
        $X = Z^TSZ$ \\
        $Z = Z(a_0I + a_1X + a_2X^2)$
        \caption{Iterative inverse overlap matrix factorization algorithm with parameter-free stopping and additional refinement procedure.}\label{algorithm:inv-overlap-wstopping-refine}
    \end{algorithm}

    \section{Numerical Results}\label{sec:numerics}

    In this section we demonstrate and evaluate the performance of the different mixed-precision inverse matrix-factorization algorithms. We will use general test examples based on synthetic overlap matrices as well as a more challenging and ill-conditioned overlap matrix, which is unfortunately not that uncommon in electronic structure calculations. All calculations were run on a local compute node with a 64-core 2 GHz AMD-EPYC 7713 Rome CPU with 4 Nvidia A100 GPUs, each with 40 GBs of memory. All GPU and Tensor core calculations were done using a single A100 GPU.
    
    \subsection{Synthetic overlap matrices}
    To explore the numerical accuracy and performance of the iterative refinement algorithm, we chose a test matrix that mimics basic properties of a dense overlap matrix: symmetry, sparsity and positive-definiteness. First, we define $\tilde{S}$:
    \begin{align}
        \tilde{S}_{i,j} := \mathrm{exp}\bigg({\frac{-|i-j|}{2}}\bigg)\sin(i+1)\;, \qquad 1\le i,j \le N\;.
    \end{align}
    To ensure positive-definiteness, we add to $\tilde{S}$ a scaled identity matrix which shifts the spectrum by slightly more than the lowest eigenvalue, $\varepsilon_1$, to get an overlap matrix that is positive definite, $S:= \tilde{S} + (\gamma - \varepsilon_1)I$. The value of $\gamma$ allows us to tune the condition number of the overlap matrix. In the examples with these synthetic matrices, we have chosen $\gamma =0.5$. The overlap $S$ is then diagonalized using the cuSolver library and its inverse square root is explicitly constructed as $S^{-1/2} = V\Lambda^{-1/2}V^T$, where $\Lambda$ are eigenvalues and $V$ are eigenvectors of $S$. Lastly, a random matrix $U$ is added to $S^{-1/2}$ that serves as the initial guess $Z_0$. Explicitly, 
    \begin{align}
        Z_0 = S^{-1/2} + \alpha U\;,
    \end{align}
    where $\alpha$ is a scaling factor and where the $U$ matrix elements, $U_{ij} = \mathcal{U}(-0.5,0.5)$, are given from a uniformly distributed random variable in the interval $[-0.5,0.5]$ with standard deviation of $1/\sqrt{12}$. The scalar $\alpha$ can then be adjusted depending upon on how far away from $S^{-1/2}$ the initial guess is desired to be. The converged inverse factor $Z$ may not be equal to $S^{-1/2}$ and the initial deviation from $S^{-1/2}$ does not depend only on $\alpha$, but also on the system size and the condition number of $S$.

    %%%%%%%%%%%%%%%%%%%%
    \begin{figure}
    \begin{tikzpicture}
        \begin{semilogyaxis}[width=.48\textwidth,height=2.5in,
                    xmin=0, xmax=9,
                    ymin=.00001, ymax=1e8,
                    legend pos=south east,
                    legend columns=1,
                    ylabel={$||X_{n+1}-I||/||X_{n}-I||^3$ },
                    xlabel={Iteration $n$},
                    scaled x ticks = false,
                    ymajorgrids=true,
                    grid style=dashed,
                    domain=100:1000,
                    legend entries={$||\cdot||_F \quad N=1024$, $||\cdot||_2 \quad N=1024$, $||\cdot||_F \quad N=16384$, $||\cdot||_2 \quad N=16384$}
                    ]
        \addplot[blue,mark=square*,mark size = 2, line width=1.5] 
                table[x index=0,y index=4, col sep=tab]{Norm1024.txt};
        \addplot[blue, dashed, mark=square*,mark size = 3, line width=1.25] 
                table[x index=0,y index=3, col sep=tab]{Norm1024.txt};
        \addplot[brown,mark=*, mark size = 2,line width=1.5]                          
                table [x index=0,y index=4, col sep=tab] {Norm16384.txt};
        \addplot[brown, dashed, mark=*, mark size = 3,line width=1.25] 
                table[x index=0,y index=3, col sep=tab] {Norm16384.txt};
        \end{semilogyaxis}
    \end{tikzpicture}
    \begin{tikzpicture}
        \begin{semilogyaxis}[width=.48\textwidth,height=2.25in,
                    xmin=0, xmax=9,
                    ymin=.0001, ymax=2,
                    legend pos=north east,
                    legend columns=1,
                    ylabel={$||X_n-I||_2$ },
                    xlabel={Iteration $n$},
                    scaled x ticks = false,
                    ymajorgrids=true,
                    grid style=dashed,
                    domain=100:1000,
                    legend entries={$N=1024$, $N=16384$ }
                    ]
        \addplot[blue,mark=square*,line width=1.5] table {Norm1024.txt};
        \addplot[brown,mark=*, mark size = 2.25,line width=1.5] table 
        {Norm16384.txt};
        \end{semilogyaxis}
    \end{tikzpicture}
    \caption{(top) Stopping ratios for the 2-norms (dotted) and Frobenius norms (solid) for matrices of size $N=1024$ and $N=16384$ using the  $(h)(h)+(h)(l)+(l)(h)$ method. (bottom) 2-norm errors for the two selected matrix sizes of $N$ = 1024 and 16384 as a function of iteration number, $n$. The 2-norm errors decay cubically for both matrix sizes and the stopping criteria is satisfied by both norms in the upper panel. $Z_0$ uses $\alpha$ = 0.005. } 
    \label{fig:stoppingThreshold}
    \end{figure}
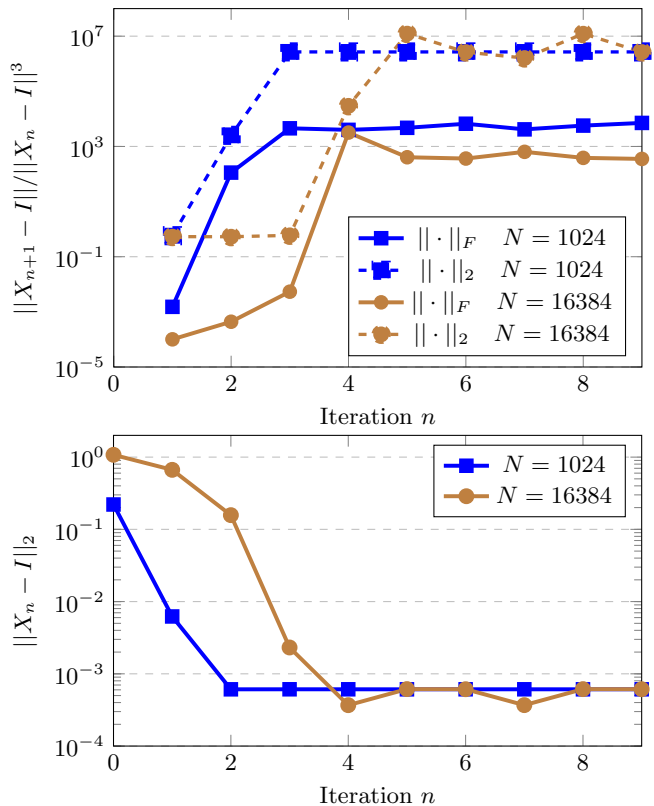
    %%%%%%%%%%%%%%%%%%%%

    \subsection{Convergence tests}
    
    Figure~\ref{fig:stoppingThreshold} displays the convergence errors (in 2-norm) in the lower panel and the convergence ratios for the stopping criteria in Eq.\ (\ref{eq:stopL2}) for the $(h)(h) + (h)(l) + (l)(h)$ method in the upper panel. The initial error scales approximately cubically, as expected.  This is observed in the bottom panel for both smaller and larger matrices with $N$ = 1024 and 16384, respectively. As expected, with a larger initial error it takes slightly longer for the algorithm to converge, which is reflected in the top panel in Fig.~\ref{fig:stoppingThreshold}, where the convergence ratio for the stopping criteria is plotted for both the 2-norm, $\|\cdot\|_2$ and the Frobenius norm, $\|\cdot \|_F$. Here, for $N$ = 1024, the convergence ratio is greater than 1 when convergence is reached at iteration number 2, as is seen in the lower panel. On the other hand, for $N$ = 16384, ratios of both norms exceed 1 at iteration number 4, as is seen in the lower panel. These tests numerically confirm the convergence condition, as derived in section~\ref{sec:stopping}. It is interesting to see the difference in the magnitude between convergence ratios in the Frobenius norm and the 2-norm. In all our numerical tests, the Frobenius matrix norms are consistently larger by about an order of magnitude compared to the matrix 2-norms and explains the shift in the convergence ratios in the upper panel. 
    %%%%%%%%%%%%%%%%%%%%
    
    \begin{figure}
    \begin{tikzpicture}
        \begin{semilogyaxis}[width=.48\textwidth,height=2.5in,
                    xmin=0, xmax=6,
                    ymin=.00001, ymax=1e10,
                    legend style={at={(0.01,0.98)},anchor=north west},
                    legend columns=1,
                    ylabel={$||X_{n+1}-I||_F/||X_{n}-I||^3_F$},
                    xlabel={Iteration $n$},
                    scaled x ticks = false,
                    ymajorgrids=true,
                    grid style=dashed,
                    domain=100:1000,
                    legend entries={$(h)(h)+(h)(l)+(l)(h)$, $(h)(h)+\bf{refi}$, $(h)(h)$, {\bf single}}
                    ]
        \addplot[black,mark=square*,mark size=2.75, line width=1.25] 
                table[x index=0,y index=4, col sep=tab]{Norm16384.txt};
        \addplot[orange, dashed, mark=square*, mark size = 2.75, line width=1.25] table [x index=0,y index=4, col sep=tab]{Norm16384-hhRefinementFP32.txt};
        \addplot[red,mark=*, mark size=2,line width=1.5] table [x index=0,y index=4, col sep=tab] {Norm16384SingleHalfFlag.txt};
        \addplot[blue,mark=o, mark size=2.75,line width=1.5] table [x index=0,y index=4, col sep=tab] {Norm16384-FP32.txt};
        \end{semilogyaxis}
    \end{tikzpicture}
    \begin{tikzpicture}
        \begin{semilogyaxis}[width=.48\textwidth,height=2.25in,
                    xmin=0, xmax=6,
                    ymin=1e-7, ymax=2,
                    legend pos=north east,
                    legend columns=1,
                    ylabel={$||X_n-I||_2$ },
                    xlabel={Iteration $n$},
                    ymajorgrids=true,
                    scaled x ticks = false,
                    ymajorgrids=true,
                    grid style=dashed,
                    domain=100:1000
                    ]
        \addplot[black,mark=square*, mark size = 2.75, line width=1.25] table {Norm16384.txt};
        \addplot[orange, dashed, mark=square*, mark size = 2.75, line width=1.25] table {Norm16384-hhRefinementFP32.txt};
        \addplot[red,mark=*, mark size = 2.00,line width=1.5] table 
        {Norm16384SingleHalfFlag.txt};
        \addplot[blue, mark=o, mark size = 2.75, line
        width=1.5] table 
        {Norm16384-FP32.txt};
        \end{semilogyaxis}
    \end{tikzpicture}
    \caption{Comparison of numerical accuracy towards convergence of four methods described in Table~\ref{tab:methods}: $(h)(h)+(h)(l)+(l)(h)$, $(h)(h)+\bf{refi}$, $(h)(h)$, and {\bf single}. (top) shows the stopping criteria using Frobenius norms. (bottom) shows errors in the 2-norm. Both the 2-norm errors and the stopping thresholds reach {\bf single} method level accuracy when we add a refinement step in single precision to $(h)(h)$ method (see blue square-marked line vs. orange square-marked dashed line). We set $N=16384$ and $\alpha$ = 0.005.}
    \label{fig:several-methods}
    \end{figure}
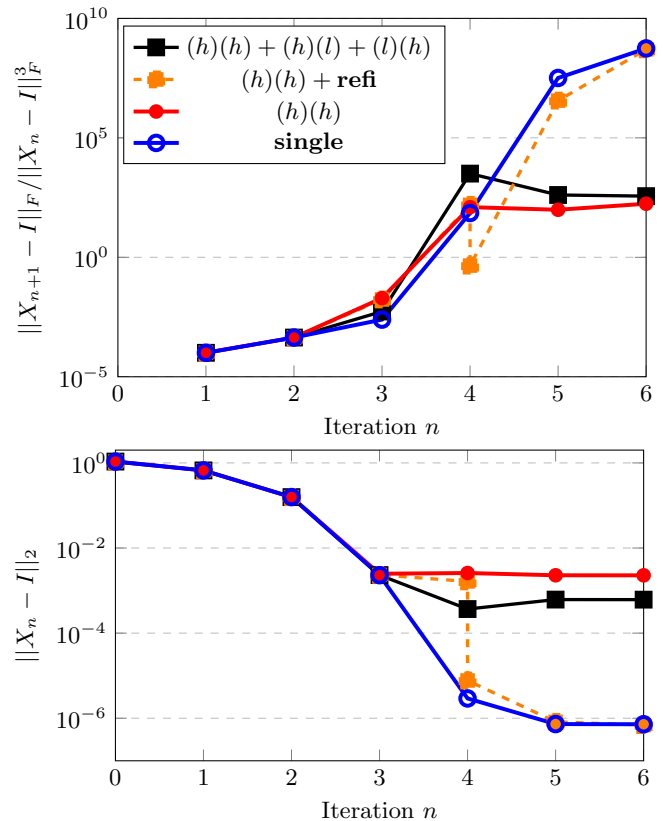
    
    \subsection{Accuracy and efficiency}
 
    Figure~\ref{fig:several-methods} shows the convergence ratios and numerical accuracy of four of our methods (see Table~\ref{tab:methods}). The bottom panel in Figure~\ref{fig:several-methods} shows how the error, $\|X_n-I\|_2$, in the initial iterations for $n = 0,1,2,3$ are almost identical for the four different methods. Only after convergence, given by the convergence ratio being $>1$ at $n = 4$ in the upper panel, do we find a divergence in the error. The computationally fastest method is the $(h)(h)$ scheme (red circles with solid line), which terminates when the error is of the magnitude 10$^{-3}$. Compared to that, the $(h)(h) + (h)(l) + (l)(h)$ method is more accurate by a factor of $\sim$5. Both methods use Tensor cores and therefore, due to round-to-zero rounding in Tensor core hardware,\cite{MFasi2023} their accuracy is expected to be less than the {\bf single} method. Including an additional refinement step to either of these Tensor-core-applicable methods, once they have converged, provides a more accurate inverse factor, $Z$, although at an added cost. The accuracy of $(h)(h)+\bf{refi}$ is shown in Fig.~\ref{fig:several-methods} (the dashed orange line) where the 2-norm error drops to the same level as that of the {\bf single} method (solid blue line). 

    The top panel of Figure~\ref{fig:several-methods} shows the stopping criteria for several methods using Frobenius norms. Here too, the ratios are similar for all the methods as the algorithm reaches convergence. At convergence, $(h)(h) + (h)(l) + (l)(h)$ has a slightly higher ratio than the $(h)(h)$, which reflects the lower accuracy of $(h)(h)$. In fact, it would not be unreasonable to think $(h)(h)$ could actually fail for more challenging
    overlap matrices, i.e. if they are ill-conditioned. Applying a single precision refinement step at the convergence reduces errors substantially and, in effect, acts like a newly restarted algorithm with a closer initial guess. The dashed orange line at iteration number
    4 shows a ratio of less than 1 but is then increased drastically
    in the subsequent iteration number 5 as the single precision iteration reaches convergence. The $(h)(h)$-algorithm stops at iteration number 4. At this point we make the refinement step. Typically we would only do one extra refinement step, although we could continue if further improvements can be achieved in the convergence.  
    
\begin{figure} 
    \begin{tikzpicture}
        \begin{loglogaxis}[width=\columnwidth,height=0.95\columnwidth,
                    xmin=100, xmax=50000,
                    ymin=0.2, ymax=20000,
                    legend style={at={(-0.02,1.23)},anchor=north west},
                    legend columns=2,
                    ylabel={Time-to-solution (ms)},
                    xlabel={Matrix size ($N$)},
                    scaled x ticks = false,
                    ymajorgrids=true,
                    grid style=dashed,
                    legend entries={cuSolver (FP64), {\bf double}, {\bf single}, $(h)(h)+(h)(l)+(l)(h)$, $(h)(h)$, $(h)(h)+\bf{refi}$ }
                    ]
        \addplot[blue,mark=square*,line width=0.5] table {cusolver-a100.txt};
        \addplot[brown,mark=x, mark size = 3,line width=0.75] table 
        {double-a100.txt};
        \addplot[red,mark=triangle*, line width=0.75] table {single-a100.txt};
        \addplot[orange,mark=*, line width=0.75] table 
        {TC-a100.txt};
        \addplot[purple,mark=pentagon*, mark size = 3.00, line width=0.75] table 
        {TC-FP32FP16Internal.txt};
        \addplot[black,mark=*, mark size = 2.00, line width=0.75] table 
        {TC-InternalRefineFP32.txt};
        \end{loglogaxis}
    \end{tikzpicture}
    \caption{Plot of run time as a function of matrix size $N$ using a single Nvidia A100 GPU and its Tensor cores. For all precisions the algorithm performs well for smaller matrices since diagonalization is much less efficient for $N\lesssim 1000$ than matrix multiplication. For $Z_0$ we choose $\alpha$ = 0.005.}
    \label{fig:runtime}
    \end{figure}
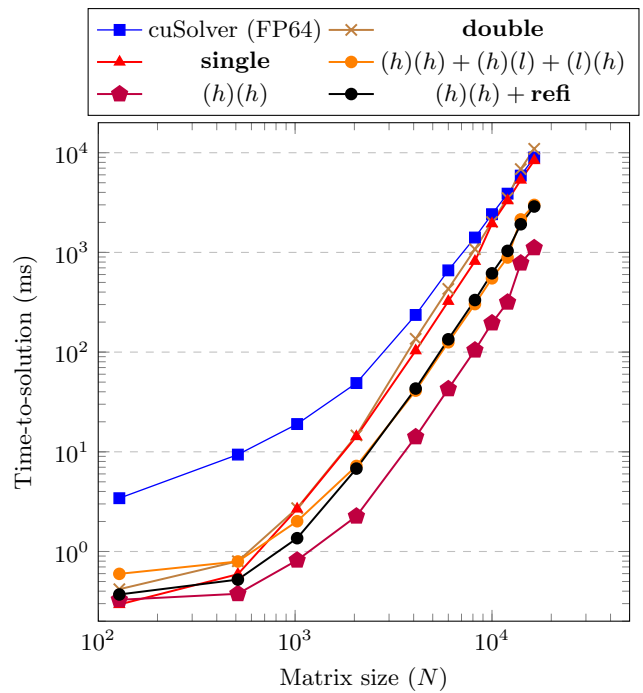

    In Figure~\ref{fig:runtime} we display wall clock times for computing the inverse overlap factor $Z$ for varying matrix sizes in different precisions. Times for a cuSolver diagonalization construction of $S^{-1/2}$ running on the GPU are also provided and can be used for direct comparison. For all precisions, in the regime of $N=\mathcal{O}(10^2)$ to $N=\mathcal{O}(10^3)$, the iterative factorization algorithm is substantially faster than constructing $S^{-1/2}$ directly through diagonalization. This is due to the lower efficiency of cuSolver and other divide-and-conquer eigensolvers on GPUs for small matrix sizes.\cite{JFattebert24} With $N=1024$, $(h)(h){\rm +{\bf refi}}$ has a wall clock time that is an order of magnitude smaller than diagonalization, and is also substantially faster than the other higher precision versions. Increasing to larger and larger matrix sizes still shows a significant advantage of $(h)(h){\rm +{\bf refi}}$, though the unfriendly scaling of the single precision refinement multiplications starts to become more apparent. For a size $N=16384$ matrix, $(h)(h){\rm +{\bf refi}}$ beats diagonalization by only a factor of about 3. Note that these times are recorded for a complete run of the factorization algorithm where the number of iterations until convergence varies as matrix sizes grow and precisions change. For example, for $N=1024$, the total number of iterations to convergence is 3 while for $N=16384$, it takes 5 iterations to converge. Moreover, the number of iterations to convergence also depends on the initial guess $Z_0$, which impacts the total solution time such that a $Z_0$ far from a real factor will increase the computational cost. Each data point in Figure~\ref{fig:runtime} is from a single calculation since only small variations in wall clock times during multiple executions were observed, and all precision implementations exhibit a smooth increase in time as a function of matrix size.
    
    One pleasant surprise, is that even on ordinary GPUs (without using Tensor cores), the iterative refinement algorithm is quite competitive with diagonalization and is substantially faster for matrix sizes of about 5000 or less. A crossover happens around $N=10^4$ where computing $S^{-1/2}$ directly with diagonalization starts to become more efficient. Nevertheless, there is a wide range of matrix sizes for which the iterative refinement method in any precision is more efficient than using diagonalization. Because of the finite interatomic range in the overlap between local basis sets, we can expect the matrices to be very sparse for the larger system sizes ($>$ 10$^4$). In these cases, a numerically thresholded sparse matrix implementation of the inverse factorization algorithm (running on a GPU or a CPU), which typically can achieve linear scaling in the computational cost as a function of the system size, is likely faster.\cite{CNegre16} Currently, sparse-sparse matrix-multiplications cannot easily be performed on Tensor cores (see Ref.~\citenum{OZachariadis20}) and for both CPUs and GPUs they introduce a significant reduction in efficiency, i.e.\ the number of floating point operations per second that can be performed.
    
    %%%%%%%%%%%%%%%%%%%%

    \subsection{Silver Nanoparticle Overlap Matrix}\label{sec:ag561}
    In quantum chemistry applications overlap matrices are sometimes very
    ill-conditioned. As an illustrative example, we take an overlap matrix built for simulation of the electronic structure of an icosahedral-shaped silver nanoparticle with 561 atoms, Ag$_{561}$ (shown in the inset of Figure~\ref{fig:normAg561} top panel) . The basis set for this system relies on atom-centered orbitals, 18 for each atom ($N$ = 10098), capturing the essence of $s$, $p$, and $d$ orbitals. These localized atomic orbitals are linearly combined to construct pseudo-wavefunctions within the projector-augmented wave functions (PAW) formalism to solve the ground state electronic energy in the open source software GPAW package. \cite{LarsenAg561BasisSet09, KuismaAg561BasisSet2015, EnkovaaraGPAWSoftware2010} 
    Figure~\ref{fig:normAg561} (top panel, blue curve), shows the eigenvalue distribution, $\varepsilon_n$, of the overlap matrix, calculated using the diagonalization routine \texttt{numpy.linalg.eig} from NumPy.\cite{HarrisNumPy2020} The overlap matrix is quite ill-conditioned with many eigenvalues close to 10$^{-5}$ and others close to $10^1$, so that the condition number is on the order of $10^6$. Additionally, towards the lower end of the spectrum, some $\varepsilon_n$ are clustered and very close, making these states almost degenerate. To perform operations on ill-conditioned matrices in low precision is a challenge and we expect that the accuracy of the inverse factorization algorithm to be significantly reduced.

    Figure~\ref{fig:normAg561} (lower panel) demonstrates how the $(h)(h)+(h)(l)+(l)(h) + {\bf refi}$ algorithm converges for the ill-conditioned overlap matrix. The 2-norm errors are shown for two initial scaling factors, low $\alpha = 0.001$ which makes $Z_0$ close to $S^{-1/2}$ and high $\alpha = 0.007$, ensuring that $Z_0$ is farther from the actual $S^{-1/2}$. When the initial guess for $Z_0$ is far from the actual inverse factor, $\|X_n-I\|_2$ is initially larger. This also implies that the algorithm will take more iterations to reach convergence. In the lower panel of Fig.\ \ref{fig:normAg561} we see that $\|X_n-I\|_2$ has initially plateaued already after the first iteration for $\alpha = 0.001$, while for $\alpha = 0.007$, initial convergence is reached when $n = 3$.  In the initial steps before the refinement, we only reach a fairly low accuracy with a convergence of about $10^{-2}$ because of the ill-conditioning. Once the initial convergence is reached (as determined by the convergence ratio being larger than 1) an additional refinement step is performed in double precision. Transforming to double precision and completing Step 2 in Sec.~\ref{sec:refinement} reduces the error to order 10$^{-6}$. After completing one additional iteration, Step 3 of Sec.~\ref{sec:refinement}, the error is around 10$^{-11}$. 

    Because of the large condition number, it is important to do the refinement step in double precision. A single precision refinement only lowers the error to around 10$^{-3}$. However, a double precision refinement step is not substantially more expensive than single precision refinement when the matrix size is in the 10$^4$ range. For the $\alpha = 0.001$ case, adding a single refinement step takes a total time-to-solution of around 1286 milliseconds, while a double precision refinement step adds a mere $\sim$16 additional milliseconds to this. This compares to a time-to-solution of 1652 ms for the {\bf double} scheme (see Table \ref{tab:methods}) and a time of 2643 ms when using the cuSolver diagonalization to construct $S^{-1/2}$.

    \subsection{Padding $Z_0$ to enhance Tensor core performance}

\begin{figure}
        \begin{tikzpicture}
            \node[inner sep=0pt] (Ag561) at (2.2,2.6)
            {\includegraphics[width=.18\textwidth]{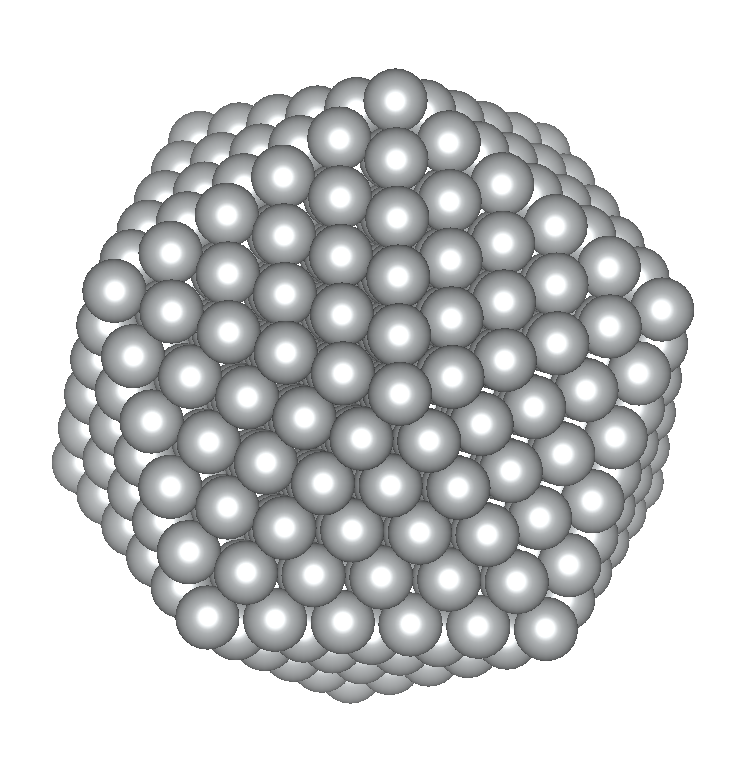}};
            \node [anchor=west] (note) at (1.5,4.25) {\large Ag$_{561}$};
            \begin{loglogaxis}[width=.48\textwidth,height=2.5in,
                                xmin=0.5, xmax=20000,
                                ymin=.00001, ymax=100,
                                legend pos=north east,
                                legend columns=1,
                                ylabel={Eigenvalue, $\varepsilon_n$},
                                xlabel={Eigenvalue \#, $n$},
                                ymajorgrids=true,
                                scaled x ticks = false,
                                ymajorgrids=true,
                                grid style=dashed
                    ]
                    \addplot[blue,mark=square*, mark size = 0.1, line width=1.0] table {overlap-eigval.txt};
            \end{loglogaxis}
        \end{tikzpicture}\label{fig:eigvalAg561}
        \begin{tikzpicture}
            \begin{semilogyaxis}[width=.48\textwidth,height=2.5in,
                    xmin=0, xmax=7,
                    ymin=1e-12, ymax=20,
                    legend pos=north east,
                    legend columns=1,
                    ylabel={$||X_{n}-I||_2$},
                    xlabel={Iteration $n$},
                    scaled x ticks = false,
                    ymajorgrids=true,
                    grid style=dashed,
                    %domain=100:1000,
                    legend entries={$\alpha$=0.001, $\alpha$=0.007}
                    ];
                \draw [dotted,->] (axis cs:4,1e-12) -- (axis cs:4,5e-7);
                \draw [dotted,->] (axis cs:2,1e-12) -- (axis cs:2,5e-7);
                \addplot[black,mark=halfcircle*,mark size=2.75, line width=1.25] 
                table[x index=0,y index=1, col sep=tab]{NormAg561-IntRefFP32-lowPert.txt};
                \addplot[orange,mark=halfdiamond*,mark size=2.75, line width=1.25] 
                table[x index=0,y index=1, col sep=tab]{NormAg561-IntRefFP32-highPert.txt}; 
            \end{semilogyaxis}
        \end{tikzpicture}
        \caption{Eigenvalues of the overlap matrix for a silver nanoparticle Ag$_{561}$ and convergence of the algorithm for this system. (Top) shows overlap eigenvalues solved using NumPy eigensolver, and the inset shows the silver nanoparticle of interest. (bottom) presents $||X_{n}-I||_2$ for two initial $Z_0$ cases where $\alpha=0.001$ (black with half filled circles) means $Z_0$ is quite close to $S^{-1/2}$ and $\alpha=0.007$ makes $Z_0$ farther from $S^{-1/2}$ (orange line with half filled diamonds). The dotted vertical lines at iteration number 2 and 4 show the actual stopping point of the algorithm (when the convergence ratio exceeds 1), before the refinement step is completed.} 
        \label{fig:normAg561}
    \end{figure}
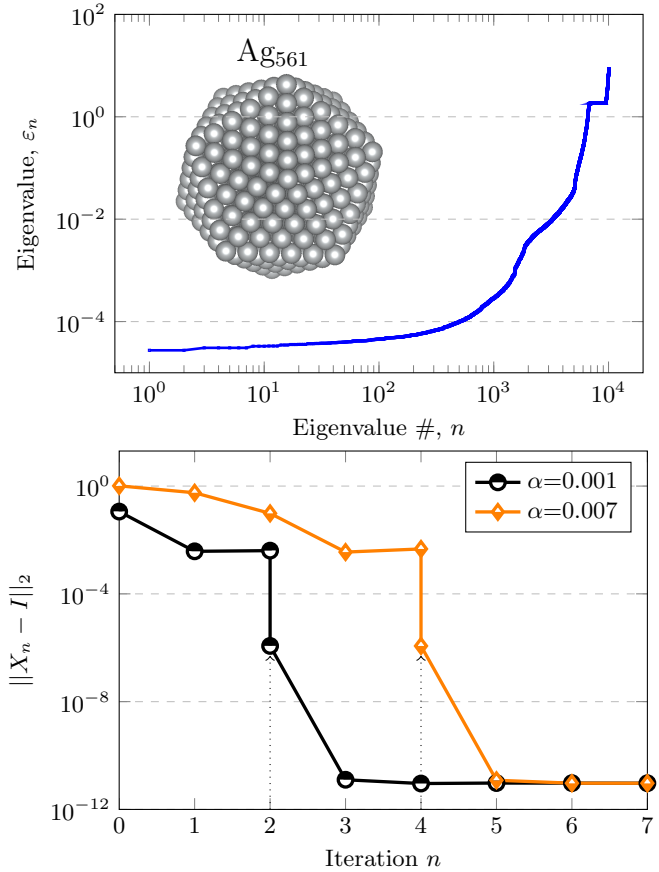

    Tensor cores require certain matrix dimensions to achieve optimal performance. Based on our numerical tests, this penalty for using non-optimal dimensions can result in a performance loss of as much as a factor of two. In section 2.1.11 of the Nvidia cuBlas documentation several conditions on matrix dimensions necessary to achieve optimal performance are discussed.\cite{cublas} One of the conditions, is the requirement that the inner matrix dimension should be divisible by $8$. Of course in many applications, such as the silver nanoparticle in the Section \ref{sec:ag561}, matrices do not always have these favorable sizes. Fortunately, this can be avoided by padding matrices with additional zeros. In the case of $\alpha=0.001$, in Fig.\ \ref{fig:normAg561}, time-to-solution for $(h)(h)+(h)(l)+(l)(h)+\bf{refi}$ improves by 35\% to 846 ms. Use of padding then yields a 3x speed up over constructing $S^{-1/2}$ with diagonalization instead of only a 2x gain without padding.

    \section{Conclusion}
    AI-optimized hardware like Nvidia's Tensor cores, provide huge boosts in speed and efficiency over ordinary GPUs, but their innate low precision makes them more challenging to use for scientific computing. This article focuses on how we can apply AI-hardware accelerators to electronic structure calculations. Our target problem is the factorization of the inverse overlap matrix required for the congruence transformation of the generalized quantum-mechanical eigenvalue problem, defined by $HC=SC\epsilon$, into a standardized one, $H^\perp C^{\perp}=C^{\perp}\epsilon$. This factorization, determined by $Z^TSZ = I$, is computed using an iterative matrix solver in mixed precision along with a novel non-parametric stopping criteria that enables efficient use of Tensor cores. Using simple, well-conditioned test cases, we show that for a broad range of overlap matrix sizes, from around a couple hundred to more than ten thousand, our technique is significantly faster than using diagonalization to construct the L{\"o}wdin orthogonalization factor $S^{-1/2}$, assuming a good initial guess. Additionally, even in the case of a realistic silver nanoparticle with an ill-conditioned overlap matrix, we achieve a more than 300\% speedup over diagonalization for a similarly accurate factor $Z$.

    In electronic structure calculations we often have access to a good initial guess of $Z$, e.g. from the previous configuration of a geometry optimization or from an incomplete or numerically thresholded approximate calculation.\cite{benzi96,benzi98,mchallacombe99,ANiklasson05b} However, of particular importance are molecular dynamics simulations. In this case we often need high performance to be able to reach relevant simulation timescales. The initial guess, $Z_0$, estimated from previous time steps \cite{ppulay04,jm_herbert05,TKuhne07} or propagated using extended Lagrangian Born-Oppenheimer molecular dynamics,\cite{ANiklasson07,ANiklasson08} is often very close to the exact solution and therefore appropriate for the formulations discussed in this paper.  

    \section{Supporting Information}
    The Supporting Information (SI) is available free of charge at {[}url{]}. The SI provides the Python script showing implementation of Algorithm~\ref{algorithm:inv-overlap-wstopping} for three cases, {\bf double}, {\bf single} and $(h)(h) + (h)(l) + (l)(h)$.
    
    \section{Acknowledgments}
    This work is supported by the U.S. Department of Energy Office of Basic Energy Sciences (FWP LANLE8AN), by the LANL LDRD-ER program, and by the U.S. Department of Energy through the Los Alamos National Laboratory. We thank the CCS-7 group and Darwin cluster at Los Alamos National Laboratory for computational resources. Darwin is funded by the Computational Systems and Software Environments (CSSE) subprogram of LANL’s ASC program (NNSA/DOE). Additionally, we also thank LANL Institutional Computing Program for providing computational resources for testing purposes.
    \newpage

    \appendix
    \section{Coefficients of stopping criteria}
    Here we show how the coefficients in Eq.~(\ref{eq:Xp1}) are derived. Given Eq.~(\ref{eq:np1}), we have
    \begin{align}
        Z_{n+1}^TSZ_{n+1} &= (a_0I + a_1X_n+a_2 X_n^2)^T \\ \nonumber
        &\mathrel{\phantom{=}} (Z^T_nSZ_n)(a_0I + a_1X_n+a_2 X_n^2).
    \end{align}
    Using the fact that $X_n=I+\delta_n$, this expression becomes
    \begin{align*}
        Z_{n+1}^TSZ_{n+1} &= \big(a_0I + a_1(I-\delta_n)+a_2 (I-\delta_n)^2\big)^T\\ \nonumber
        &\mathrel{\phantom{=}}(I-\delta_n)\big(a_0I + a_1(I-\delta_n)+a_2 (I-\delta_n)^2\big)\;.
    \end{align*}
    Expanding these expressions and collecting terms with common factors leads to
    \begin{align}
        Z_{n+1}^TSZ_{n+1} &= \bigg(a_0^2 + 2a_0a_1 + 2a_0a_2 \nonumber + a_1^2 + 2a_1a_2 + a_2^2 \bigg)\delta_0^0\\ \nonumber
         +{} &\bigg(a_0^2 + 4a_0a_1 + 6a_0a_2 +3a_1^2 + 8a_1a_2 + 5a_2^2 \bigg)\delta_0^1\\ \nonumber
        +{} &\bigg(2a_0a_1 + 6a_0a_2 + 6a_1^2 + 12a_1a_2 + 10a_2^2  \bigg)\delta_0^2\\ \nonumber
        +{} &\bigg(2a_0a_2 + a_1^2 + 8a_1a_2 + 10a_2^2  \bigg)\delta_0^3\\  \nonumber
        +{} &\bigg(2a_1a_2 + 5a_2^2 \bigg)\delta_0^4 \\ 
        +{} & a_2^2\delta_0^5.
    \end{align}
    The values of $a_k$ are taken from Ref.~\citenum{ANiklasson04b}'s Table I, and are $a_0$ = 15/8, $a_1$ =-5/4 and $a_2$ = 3/8. Plugging these in, we calculate the coefficients for each power of $\delta_n$. The coefficient for the zeroth-order term is 1 while coefficients of $\delta_n$ and $\delta_n^2$ vanish. Hence, 
    \begin{align}
        c &= \bigg(2a_0a_2 + a_1^2 + 8a_1a_2 + 10a_2^2 \bigg) \\ \nonumber
        &= 2\cdot\frac{15}{8}\cdot\frac{3}{8} + \bigg(\frac{-5}{4}\bigg)^2 + 8\cdot\frac{-5}{4}\cdot\frac{3}{8} + 10\bigg(\frac{3}{8}\bigg)^2  \\ \nonumber
        &= \frac{90}{4} + \frac{25}{16} + \frac{-120}{32} + \frac{90}{64} \\ \nonumber
        &= \frac{180+100-240}{64} \\ \nonumber
        &= \frac{5}{8}.
    \end{align}
    Similarly, we find the coefficient of $\delta_n^4$ to be
    \begin{align}
        d &= \bigg(2a_1a_2 + 5a_2^2 \bigg) \\ \nonumber
        &= 2 \cdot \frac{-5}{4}\cdot \frac{3}{8} + 5\cdot \bigg(\frac{3}{8}\bigg)^2 \\ \nonumber
        &= \frac{-60+45}{64} \\ \nonumber
        &= \frac{-15}{64}.
    \end{align}
    As for $e = a_2^2$, we get $(\frac{3}{8})^2 = \frac{9}{64}$. The sum of absolute values of these non-zero coefficients determines the stopping condition in Eq.~(\ref{eq:stopL2}).
   
    \bibliographystyle{achemso.bst}
    \bibliography{bib}

    \end{document}